\documentclass{cccg19}
\usepackage{graphicx,amssymb,amsmath,url,latexsym}

\usepackage{xcolor}

\usepackage{pifont}
\newcommand{\cmark}{\ding{51}}%
\newcommand{\xmark}{\ding{55}}%

\newcommand\red[1]{\textcolor{red}{#1}}
\newcommand\blue[1]{\textcolor{blue}{#1}}

\newcommand{\squeezelist}{\setlength{\itemsep}{0pt}}

\title{Unfolding Polyhedra}

\author{Joseph O'Rourke\thanks{
        Smith College, {\tt jorourke@smith.edu}
        }}

\index{O'Rourke, Joseph}

\begin{document}
\thispagestyle{empty}
\maketitle

\begin{abstract}
Starting with the unsolved ``D{\"u}rer's problem" of edge-unfolding a convex polyhedron
to a net, we specialize and generalize 
(a) the types of cuts permitted, and (b)~the polyhedra shapes,
to highlight both advances established and which problems remain open.
\end{abstract}

\section{Introduction}
\emph{D{\"u}rer's problem} asks whether every convex polyhedron 
may be cut along edges and unfolded to a single non-overlapping simple polygon in the plane,
a \emph{net}~\cite{do-gfalop-07}~\cite{o-dp-13}.
This is attributed to D{\"u}rer because he drew many such unfoldings ca.~1500,
although the question was not formulated mathematically until 1975~\cite{s-cpcn-75}.
It remains open, although there has been recent (minor) progress~\cite{o-eunfcc-17}~\cite{o2017addendum}.
Here we survey several generalizations and specializations of this central problem,
emphasizing what is settled and what remains unresolved.

Unfolding the surface of a polyhedron to a single, flat piece in the plane requires that
the cuts form a spanning tree of the vertices.
We classify cuts in four types $\cal{C}$:
\begin{enumerate}
\squeezelist
\item \emph{edge-unfold}: All cuts are polyhedron edges, as in D{\"u}rer's problem.
\item \emph{anycut-unfold}: The cuts may be generalized to any curve on the surface that form a spanning tree of the 
vertices.\footnote{
``Anycut'' is new terminology, intended to replace the ``general unfoldings'' in~\cite{do-gfalop-07}.}
\item \emph{edge-unzip}: The cut edges form a Hamiltonian path of the $1$-skeleton. This natural specialization was
introduced by Shephard~\cite{s-cpcn-75} and explored as ``unzipping" in~\cite{lddss-zupc-10}.\footnote{
``Unzipping" is my slight variation on their ``zipper unfoldings."}
Most classes of polyhedra do not admit edge-unzippings~\cite{ddeo-polycubes-2019}.
\item \emph{anycut-unzip}: The cuts form a simple curve on the surface that includes every vertex.
So a generalization (anycut) of a specialization (unzipping).
\end{enumerate}

The second classification we explore varies the shapes $\cal{P}$ of the polyhedra:
\begin{enumerate}
\squeezelist
\item \emph{convex polyhedra}: All faces convex, all dihedral angles ${\le}\pi$, as in D{\"u}rer's problem.
\item \emph{spherical polyhedra}: Specializing that all vertices lie on a sphere~\cite{o2015spiral}.\footnote{
``Spherical'' is often called \emph{inscribed} in the literature, and ``spherical polyhedra'' are
sometimes tilings of the sphere by geodesic arcs.}
\item \emph{nonconvex polyhedra}. A broad generalization, and where most applications lie.
\item \emph{orthogonal polyhedra} form an important subclass
of nonconvex polyhedra~\cite{bddloorw-uscop-98} \cite{o-uop-08} \cite{damian2007epsilon} \cite{damian2017unfolding}.
All faces meet at right angles.
\item \emph{polycubes}: Polyhedra built by gluing unit cubes whole-face to whole-face.
Here all cube edges, even those with dihedral angle $\pi$, are available for cutting.
So these are potentially easier to edge-unfold than are orthogonal polyhedra~\cite{RichaumeAndres}.
\end{enumerate}

For each class of polyhedra $\cal{P}$, and each type of cuts $\cal{C}$, we can ask:
\begin{center}
\fbox{\fbox{%
Can every polyhedron in  $\cal{P}$ be $\cal{C}$-unfolded to a net?%
}}
\end{center}
The status of these $4 \times 5 = 20$ questions is summarized in Table~1:
$6$ are unresolved.

\begin{table*}[h]
\centering 
\begin{tabular}{| l | c | c | c | c |}
\hline
\mbox{} & \mbox{} & (\emph{Specialize}) & (\emph{Generalize}) & (\emph{Gen/Spec}) \\
\emph{Shapes} & \emph{Edge-Unf} & \emph{Edge-Unzip} & \emph{Anycut-Unf} & \emph{Anycut-Unzip}\\
\hline
\hline
Convex Polyh & \red{\textbf{?}} & \xmark & \blue{\cmark} & \red{\textbf{?}}  \\
\hline
\hspace{3mm}Spherical  & \red{\textbf{?}} & \xmark  & \blue{\cmark} & \red{\textbf{?}}  \\
\hline
Nonconvex Polyh & \xmark & \xmark  & \red{\textbf{?}}  & \xmark \\
\hline
\hspace{3mm}Orthogonal & \xmark & \xmark  & \blue{\cmark} & \blue{\cmark} \\
\hline
\hspace{6mm}Polycubes & \red{\textbf{?}} & \xmark   & \blue{\cmark} & \blue{\cmark} \\
\hline
\hline
\end{tabular}
\caption{Open Problems: \red{\textbf{?}}=open, \blue{\cmark}=proven true, \xmark=counterexamples. }
\end{table*}


\small
\bibliographystyle{alpha}
\bibliography{Abstract}

%
%
%

\end{document}